\documentclass[conference, letterpaper]{IEEEtran}
\ifCLASSINFOpdf
\else
\fi
\ifCLASSINFOpdf
   \usepackage[pdftex]{graphicx}
\else
\fi

\usepackage[cmex10]{amsmath}
\usepackage{color}
\usepackage{fancyhdr}
\usepackage[caption=false,font=footnotesize]{subfig}
\usepackage{hyperref}

\renewcommand{\thispagestyle}[2]{} 

\fancypagestyle{plain}{
        \fancyhead{}
        \fancyhead[C]{first page center header}
        \fancyfoot{}
        \fancyfoot[C]{first page center footer}
}
\begin{document}

%
\title{A blockchain-based Decentralized System for proper handling of temporary Employment contracts}


\author{\IEEEauthorblockN{Andrea Pinna, Simona Ibba}
\IEEEauthorblockA{Department of Electrical and Electronic Engineering (DIEE)\\
University of Cagliari\\
Piazza D’    Armi, 09100 Cagliari, Italy.\\
Email: a.pinna@diee.unica.it, simona.ibba@diee.unica.it}
}


%


\maketitle

\begin{abstract}

Temporary work is an employment situation useful and suitable in all occasions in which business needs to adjust more easily and quickly to workload fluctuations or maintain staffing flexibility. Temporary workers play therefore an important role in many companies, but this kind of activity is subject to a special form of legal protections and many aspects and risks must be taken into account both employers and employees.  In this work we propose a blockchain-based system that aims to ensure respect for the rights for all actors involved in a temporary employment, in order to provide employees with the fair and legal remuneration (including taxes) of work performances and a 
protection in the case employer becomes insolvent. At the same time, our system wants to assist the employer in processing contracts with a fully automated and fast procedure. To resolve these problems we propose the D-ES (Decentralized Employment System). We first model the employment relationship as a state system. Then we describe the enabling technology that makes us able to realize the D-ES. In facts, we propose the implementation of a DLT (Decentralized Ledger Technology) based system, consisting in a blockchain system and of a web-based environment. Thanks the decentralized application platforms that makes us able to develop smart contracts, 
we define a discrete event control system that works inside the blockchain. 
In addition, we discuss the temporary work in agriculture as a interesting case of study.  

\end{abstract}



\begin{IEEEkeywords}
blockchain, temporary employment, smart contracts, discrete event model
\end{IEEEkeywords}

%
\IEEEpeerreviewmaketitle

\section{Introduction}
Temporary work contracts play a critical role in the current world economic and social context. Increasing international competition, slow economic growth and high unemployment rates have lead to the creation of greater job flexibility in many countries and institutions. The diffusion of non-standard contractual arrangements is also largely facilitated by technological innovations. In a global economic context, the competitiveness of companies is closely linked to the ability to adapt rapidly to new challenges and changes.
Temporary employment could be an important and flexible business tool, in order to react to the market fluctuations, affected by economic policies and some seasonal conditions. As a result, according to International Labour Organization  \cite{ILO2016}  atypical contractual arrangements are a feature of the contemporary world of work. 

However non-standard work contracts are often characterized by a lack of workers' guarantees, insecurity, low wages, limited growth prospects, lack of vocational training and less access to social security systems. 
Moreover young people, regardless of the level of education they have and the skills they possess, are engaged in non-standard jobs more frequently than other groups of the population.

In addition to this labor flexibility is also often associated with the absence of a development model combining the social quality and sustainability of new forms of work to economic growth and enabling transition from one job to another and more generally, a full acquisition of human rights.
It is therefore necessary to build social protection instruments that do not only protect workers from the risk, but also from the charm of flexibility. 


In our work we want to show how blockchain technology can be used to address the fundamental problems that occur around temporary employment, in order to protect employees and to prevent that the competition being distorted in the benefit of those companies that wish grow on the backs of exploited illegal workers. 

Both businesses and employees need the recognition of the value of their work, and, in this complex scenario of non-standard work contracts, the use of  blockchain technology of may be an excellent solution in order to ensure reliability, transparency and security. Blockchain technology indeed is based on a decentralized technical database to efficiently manage transactions. It stores these transactions in a Peer-to-Peer network. Blockchain technology is also a public registry: transactions 
consist of encrypted data that are verified and approved by the nodes participating in the network, and, subsequently, added in a block and recorded in the blockchain. The blockchain is shared between all nodes of the network. The same information are present on all nodes and therefore becomes unmodifiable unless through an operation that requires the approval of the majority of the nodes in the network. In any case, it will not change the history of these same information.
Therefore this technology introduces a new level of transparency and efficiency, by allowing the network to manage the transactions and creating confident transactions in an untrustworthy environment. 

Blockchain technology allows to quickly register work contracts with full protection of the rights both of the worker and of the employer,  in compliance with the legislation. Smart contracts are immutably saved on the blockchain and can be observed and checked for compliance at any time by the competent authorities. 

The remains of paper is structured as follows. 
Section II presents the related works. Section III describes the proposed system, the methodology that we have taken and gives the reasons why we chose the blockchain technology. In section IV we show an application of our methodology and in section V we discuss about this. Finally, in section VI we present the conclusions and some future implementations.

\section{Related works}
Blockchain technology can be used in all contexts where a decentralized system is necessary in order to ensure the involvement of many actors in the same network and guarantee a full transparency and reliability between people who do not know each other. Therefore blockchain technology is not only useful for creating digital currencies or new financial technologies, but can be applied for a wide variety of applications, such as protection systems of digital identity, provenance of documents, organizational data management, digital and physical assets.
An important research is that carried out by \cite{Faioli2016}. They, designing an interdisciplinary approach, analyze legal aspects and consequences of the use of blockchain for job organizations that want to challenge the law and the labor market.

\cite{Norta2017} instead examines the use of smart contracts, combined with intelligent multi-agent systems and Internet-of-Things devices, in order to deliver self-aware contracts with a high degree of automation for peer-to-peer collaborations. They apply a smart contract, mapped onto an automated protocol, for initiating and terminating a rental contract.

An innovative regulation of labour hours and the associated payments, is proposed by \cite{Chrono2016}. They apply recent advancements in blockchain and cryptographic technologies in order to develop a non-volatile and inflationary resistant digital asset. The transfer system considers the average hourly rate of human labour as the most fundamental unit of economic value.

The Blockchain Research Lab \cite{Steinmetz2017} developed a prototype to manage a smart contract between agency, manufacturer and worker. They focus on the case of the necessity to create a valid contract between the agency and the worker considering that the agency needs a leasing permission and the manufacturer needs enough funds to pay the agency.
The system, through a smart contract, checks the coexistence of all these requirements.

The blockchain technology is also a promising technology for the implementation of several typology of decentralized systems. In particular, in the field of a public and collaborative smart city system, blockchain represents a smart solution. In \cite{ibba2017}{ibba2017} thanks to the use of smart contracts and an Agile Involvement of Citizens, is proposed a shared and public database of eviormental signals. In \cite{mannaro2017} and \cite{mihaylov2014} the blockchain is used to implement a smart, decentralized, and free energy market. The proposed system aims to promote the installation of renewable energy domestic plants.

\section{The proposal} 
In this section we describe the technical aspect of the proposed Decentralized Employment System (D-ES). 

The proposed system is a solution that makes transparent and traceable any employment relationship, established for the temporary work. 
The system simplifies the recruitment procedure, and it is a useful tool to prevent the black labour.

In order to model the system, we first identify the actors which will be considered in the development. The model considers four typology of actors.
Two typology of actors represent human users. The first is the Employer, who creates the work activity and announces the availability of vacant posts. The second is the Worker, who applies for a temporary job.  

The other two typology of actor are components of the system. 
Because the technical difficulty of putting directly the hands in the blockchain, the DES will provide a simplified user interface, in the form of web platform. The Platform is the actor that makes users (Employers and Workers) able to create new works, to apply for them, and to access further information about the employment relationship.
And finally, the last actor is the Blockchain.  
The system will be based on the blockchain technology because its capability to provide trust and security, and for the possibility of developing decentralized application we will exploit to implement the D-ES. In the system, the blockchain has the double role: the role of ledger, public and unchangeable, and the role of control system which safeguards workers and prevents scams. In fact, thanks the availability of a decentralized virtual machine, the blockchain is not only a database but also a computing resource.

In the remainder of  this section the whole model will be discussed. In particular, at first, an introduction of the blockchain technology will be provided. Then, the technical aspects will be exposed.

\subsection{Blockchain: overview}

In order to implement the D-ES, we will use the blockchain technology. In the following we will introduce the blockchain technology and will explain the motivation of our choice.



A blockchain is a decentralized data structure, reachable and shared by the nodes of a peer-to-peer network. From a high level point of view, the blockchain (also called Decentralized Ledger Technology, DLT) can be seen as a ledger in which several typology of transaction are stored. 
Basically, with the blockchain technology a transaction is an information package in which specific data are recorded. These data include a sender and a recipient, a monetary amount, and some additional properties regarding the exchange. 

The blockchain is composed by blocks. Each block is a list of validated transactions, according with a consensus algorithm. Blocks are chained thorough a cryptographic code, called hash, computed basing on the code of the previous block and the overall digital information that compose each new block. These properties make the blockchain data unchangeable, because just a little modification will make the cryptographic codes inconsistent with the consensus algorithm. Senders and recipients of transactions have to be accounts of the blockchain, defined by an alphanumeric code called address. 

Second generation blockchains (e.g. the Ethereum blockchain) provide a computational environment, programmable by means the developing of decentralized applications called smart contracts (SC). Each SC can store data and execute function, and functions can be called by means an appropriate transaction addressed to the SC. In other words, a blockchain is a state machine, in which the state transition function is represented by the execution of transactions and SC. A new state depends on the data recorded in the previous state, composed of the information stored in the blockchain.\\

One of the most important aspect related to the blockchain is the possibility of move value or write a contract that involves a group of people, in the absence of a regulator organism. The point is that the blockchain creates the necessary trust, by means the consensus algorithm which rules the peer-to-peer network. In a fair employment relationship, employer and workers need to guarantee honesty each other, and the formation of mutual trust. Actually, if this were always true, people couldn't need of regulator organisms. 
The D-ES will use the blockchain technology in order to make clear, transparent and trusty the employment relationship between people who, at least initially, don't need to trust each other. 

The blockchain will be used to implement a transparent work ledger, and to implement the model of employment relationship described in the following.

\subsection{The D-ES state system}\label{sec:state}

In this proposal, we model an employment relationship as a state system in which states describes the current phase of the relationship. 

Taking in account the need of a legal recognition and authorization, we suppose actors of the system able to create legal employment relationships. For instance, the Employer should have the legal rights to hire people. At the same time, a Worker should be legally recognized to be able for a specific job. We represent this point defining a background system representing a generic \textit{Central Authority}. The Central Authority, in order to supervise the employment relationships, could take advantage of the transparency property of the blockchain. In the section \ref{sec:discussion} provide further details of this aspect.

\begin{figure}
\centering
\includegraphics[width=1\linewidth]{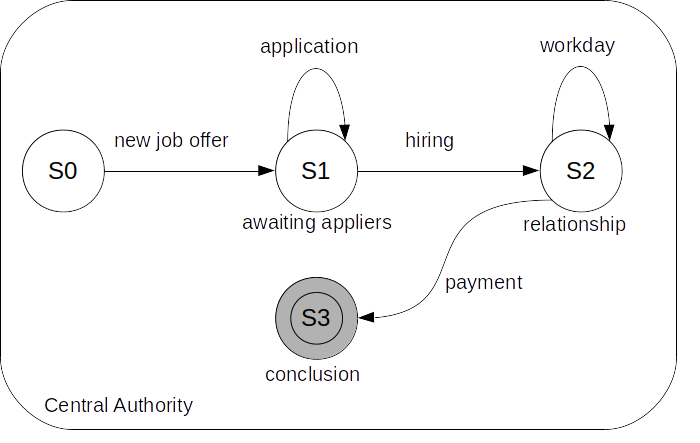}

\caption{The state diagram describing the employment phases controlled by the D-ES.}
\label{fig:statediagram}
\end{figure}

In order to describe the model, we first specify states, the events that change the state, and the role of the actors. Fig.\ref{fig:statediagram} shows the D-ES state diagram.
The initial state, S0, is an idle state. The D-ES is ready for the creation of a new employment relationship.

The first event is called “\textit{new job offer}” and it consists in the creation of a new job offer. It happens when a Employer completes the procedure to setting up a job offer, aided by the platform. The Employer uses the interface provided by the platform in order to specify all the property of the new open position: the number of working hours, the time wage, the job title, etc. 
In addition, the Employer deposits the amount of digital asset in order to cover the value of the wage. When all is done, the platform create and sends to the Blokchain some ready-made message. That messages include all the information required to create the set of smart contracts with which the D-ES will control the employment relationship. Technical detail of smart contracts will be discussed in the following.

The “\textit{new job offer}” event change the state from the initial state to the state “\textit{awaiting appliers}”. In this state, the D-ES is configured to accept the application of new workers. Now, Workers can apply for the open position. In this state, the platform shows the job offer to the workers. A internal event “\textit{application}” describes when a worker applies for the job offer. The worker has to send a message to the blokchcain, or precisely, to a smart contract. That smart contract is charged to acquire the application request, and to compute and return to the Worker, an applicant’s identification code. 

When an applicant Worker meets the Employer, they can give rise the “\textit{hiring}” event.  
The Employer now being in posses of the applicant identifier.  He sends a message to the blockchain in order to start the employment relationship.  

Now the system is arrived to the “\textit{relationship}” state. In this state, the Worker can check, at any time, its working situation and verify the number of the matured work hours. An internal event “\textit{workday}” describes the maturation of a daily number of work hours, and occurs when the Employer sends a message to the \textit{blockchain} in order to certify that the worker has completed a work day.

Automatically, when the stipulated working hours are over, the contract declares the end of the relationship and occurs the “\textit{payment}" event. During this event, the system moves the wage deposited by the Employer in the account of the Worker. This event moves from the “\textit{relationship}” to the “\textit{conclusion}” state.

\subsection{Implementation of the decentralized system}

In our proposal, the blockchain has an active role and it is a real actor of the system. Summarizing, the blockchain will identify the Employers, identify the Workers, record every employment relationship, control and compute the evolution of the employment relationship, and, finally, compute and transfer the wage from Employers to Workers. 

All these actions will be done by means the execution of a decentralized computer programs called smart contracts. The D-ES works through an decentralized ecosystem of three typology of smart contract. They are named: “sc\_deposit”, “sc\_application”, “sc\_relationship”. In order to automate the creation of new employment relationship, the three typology of smart contract will be recorded ready-made in the Platform system.

According with the state system described before, the Platform customizes the three smart contracts with the job description data provided by the Employer. At the event “new job offer”, the three smart contracts are created, configured and written inside the blockchain. Each one of the three smart contracts knows the address of the other two. 

The contract type “sc\_deposit” implements a token deposit. Token and coin deposits are a popular application in the Ethereum system. In our system, the sc\_deposit  includes a \textit{payment} function programmed to transfer the deposited wage to a specified Worker's address. This function can be called only by the sc\_relationship contract which address is recorded in the sc\_deposit memory.

Each contract type “sc\_application” provides the \textit{application} function. This function is programmed to be called by Workers' addresses and returns a cryptographic identification code, valid only for the employment relationship that it represents. In addition, that smart contract records and provides information about the job offer.

Finally, a contract type “sc\_relationship” provides the \textit{hiring} function. This function is programmed to create an employment relationship. It receives messages from the Employer address, containing the blockchain address of the Worker and his identification code as produced by the associated “sc\_application”. The contract checks the validity of the identification code.
In addition, the contract stores the current number of work hours matured by Workers. This contract is also able to ask “sc\_application” for the agreed number of work hours.
A second function \textit{workday} updates of the number of work hours when receives the appropriate message from the Employer. This function automatically computes the end of the employment relationship. In that case, in order to call the \textit{payment} function, it sends a message to the “sc\_deposit”, specifying the address of the Worker.

\subsection{The role of the Platform}
In this proposal, we anticipate the need for a Platform (that can be seen as an user interface), responsive and easy to use. This Platform provides forms and instructions that makes simple the creation of the blokchain system and the control the employment relationship. The Platform creates the three smart contract. It simplifies the creation of a new job offer and provide a friendly interface for the applicant workers. Furthermore it provides the visualization of the state of the employment relationship. 

\section{D-ES in farming: a case study}
Temporary work can be defined as an employment relationship or an agreements between two or more parties that are not explicitly governed by civil law. Their contracts are created between the parties on the basis of mutual needs that emerge during the negotiation phase. Our system want to ensure respect for the rights for all actors involved in these employment relationship.
In the agricultural sector, the problem of temporary work is greater than others sectors and in some countries, such as in Italy, the rules of occasional work in this context are different from those set out in the other production sectors.

Given that the protection of workers in the agricultural sector is a relevant social issue we decided to focus on this aspect as main use case.

In fact, the exploitation of labour in agriculture offends one of the most important activity for humanity.
Criminal organizations exploit the work force of the weakest members of society, like irregular immigrants, and acts as temporary workers' agencies, for the benefit of industrial agriculture. 
Illicit brokering and exploiters of labour are punished with the jail, according to the Law. Order forces hardly find criminals acting the exploitation of labour.

With the term \textit{caporalato} it is identified a illegal activity that concern the agricultural work. That consists in a organized form of the exploitation of labor.

In this case study we face such problem proposing a model to prevent the black labor. In particular, this D-ES application works by means a traceability system that obliges farm employers to make transparent all the working activity, starting from the employment phase and arriving to the payment phase. In addition, in this particular application, the goods produced by the work will be considered legal only if the safeguarding system produces the certification of saleability. 
Just like for the general case, in order to record and control the procedure we will use the innovative computer technology called blockchain. 

\begin{figure}
\centering
\includegraphics[width=1\linewidth]{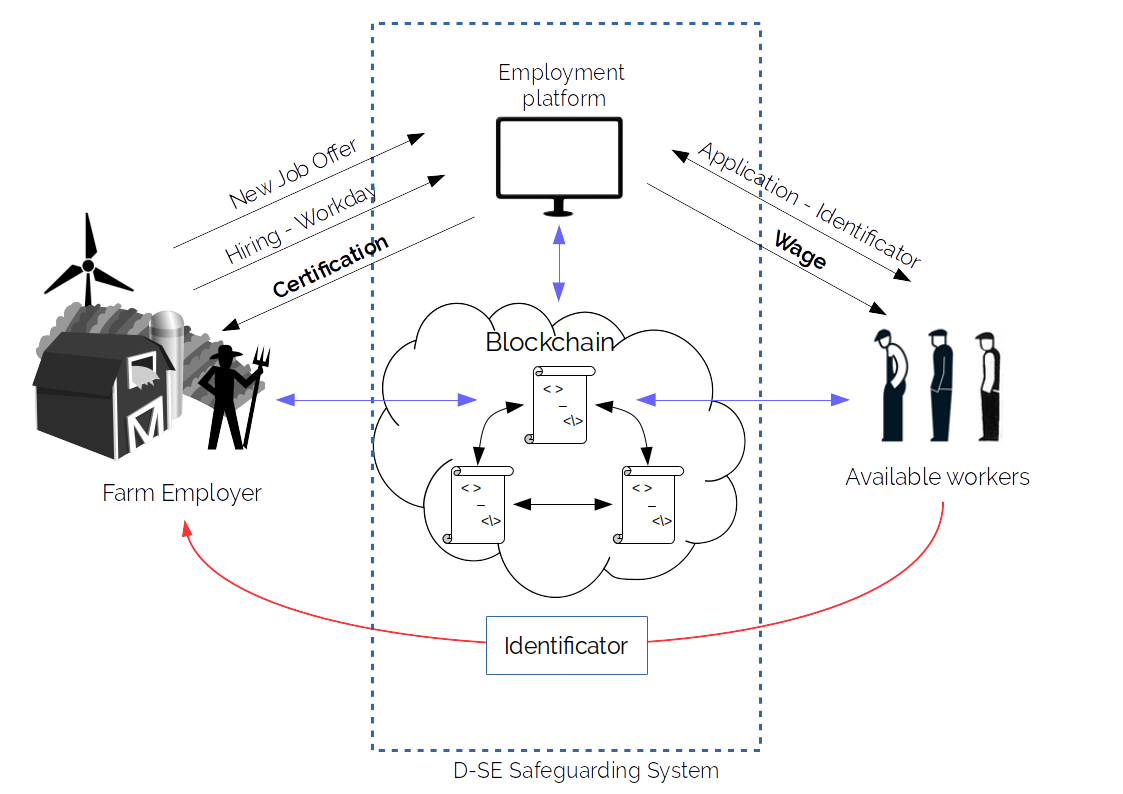}

\caption{A representation of the D-ES applied to farming temporary work. The Farm Employer create a new job offer and hires workers aided by the employment platform. The platform creates the blockchain based decentralized system which control the employment relationship. In this case study, the system not only automatically pay the salary to the workers, but provide the saleability certification of the agricultural products.
}
\label{fig:safeguarding1}
\end{figure}

\subsection{The D-ES discrete event model in farming}


In this case study we consider that a farming activity could need several workers. For instance, traditional harvesting of vegetables, involves generally  several temporary farmers. For this reason, the sc\_relationship will be configured to accept more than one Worker.

In order to describe the model of the case of study, the following simplifying assumptions was made:

\begin{itemize}
	\item Farmers and Employer are legally enabled by a central authority (i.e. an authorized employment office)
	\item exists only one typology of job and wage.
	\item Farmer always completes the job,
	\item works can be described with the number of work hours
\end{itemize}
\begin{figure*}[ht]
\centering
\includegraphics[width=0.8\linewidth]{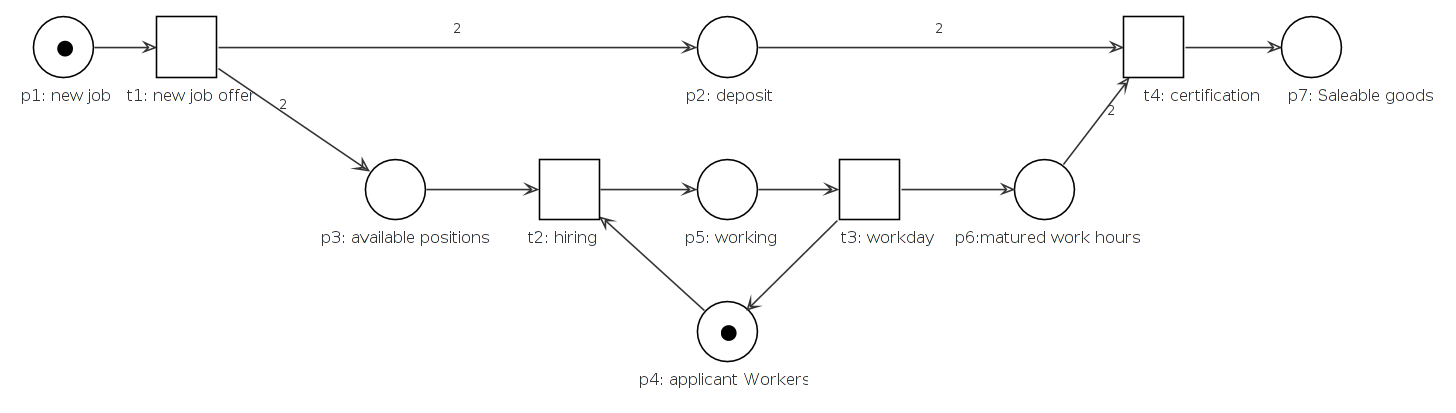}
\caption{Petri net of the safeguarding system. The marking M(4) of the place p4 represents the number of Workers who apply for the job offer. In this case, M(4)=1 or rather only one worker applies for the job. The weight of the arcs represents the number of work hours expected for the job. In the represented case the weight is equal to 2.}
\label{fig:safeguarding}
\end{figure*}



The D-ES state system described in the section \ref{sec:state} stills valid. In this case, a  
In order to obtain a high-level description of the dynamic evolution of the employment relationship, we provide a Petri Net model in which the state of the system is described by the marking. Squares represents the transitions and circle the places. Places can hold the marking. When a transition fire, the marking move from the precedent places to the following places, according the weight and the direction of the arcs that connect the two typology of nodes.

The Petri net conclude the evolution when no more transition can fire. In our case, the last transition represents the certification process. 
In the following we report the transitions description.
T1: new job offer. The safeguarding system elaborated the request of a new farming work. Basing on the job description (i.e. the number of work hours, and the expected work product), the system created the three contracts described in section \ref{sec:state} and collect the wage deposit. 

T2: hiring. An applicant Worker applies meet the Employer and been hired. 

T3: workday. A worker finish a work day and the contract sc\_relationship update the amount of matured work hours.

T4: certification. This final transition is peculiar of the farming case study. Whit the firing of this transition, the system represent the occurrence of the conditions that certificate the conclusion of the work, the correct conclusion of the all temporary employment relationship and the transfer of the wage to farmers accounts. In addition, the firing of T4 create the cryptographic certification code with which the Employer can sell its farming goods to the market.

Places description.

P1: new job. Token in this place are the representation of a work description. In this proposal, each token describes two feature, namely the expected work result (i.e the approximate quantity of goods) and the expected need of men hours. In this simplified model, P1 is marked with only one token. 

P2: deposit. This place represent the existence of a deposited wage in the sc\_deposit contract. The marking could represent the number of daily wage deposited, according with the specification defined in the creation event of the new job offer. 

P3: available positions. When a marking is present, at least one open position is available. The marking value represents the total number of work day of the open positions.

P4: applicant Workers. This place marking represents the number of workers that could apply for the job offering. 

p5: working. This place represents the number of farmers currently hired by the Employer.

P6: matured work hours. The marking in this place represents the total number of matured work hours by the hired workers.

P7: salable goods. This place represents the correct end of all employment relationship and of the farming activity and the creation of a legal certification of saleability. 

The initial marking is equal to zero in all places except places P1 and P4. The marking in P1 represents the will to create a new job offer. The marking $M$ in P4 represents the number of farmer able to apply for the job offer. In Fig. \ref{fig:safeguarding} we represents the Petri Net of the D-ES in farming considering the case of only one applicant Worker $M(P4)=1$. Note that the weight of some arcs is equal to n=2. With n we represent the number of workdays (or the number of work hours) of the job offer. 

\begin{figure}
\centering
\includegraphics[width=0.8\linewidth]{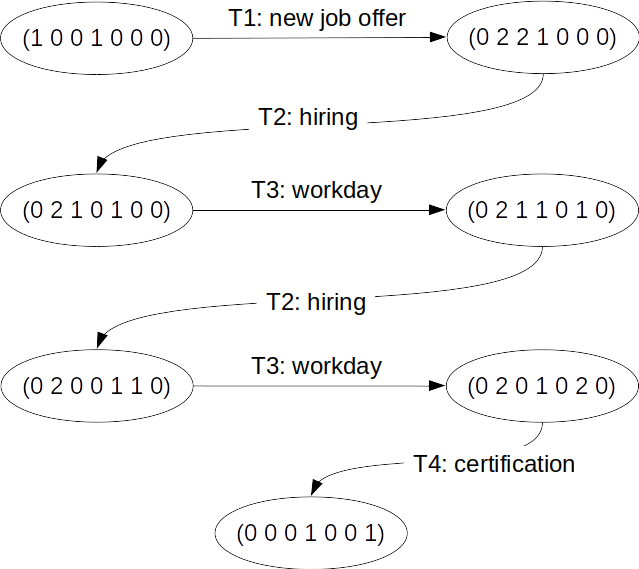}

\caption{The reachability graph of the Farming Employment relationship Petri Net in Fig\ref{fig:safeguarding}
}
\label{fig:reach}
\end{figure}

The reachability graph shown in Fig. \ref{fig:reach} of this Petri Net has seven states. In particular, the last state is characterized by a marking in the place seven M(7)=1 and in the place four M(4)=1. The last state is reachable only after the firing of T4, as described above. 

\section{Discussion}\label{sec:discussion}

Traceability is one of the most promising application of the blockchain technology. 
Our system provides high-performance in managing temporary contracts by addressing some of the key aspects that make them unsuitable for being applied to the the present working context. It aims to protect the rights of workers and enterprises at the same time, but also to ensure full control of the competent authorities during the verification of the necessary requirements for signing of the contract (the conditions related to employer and worker must be  fulfilled at the same time) and during the verification of proper performance of the contract.
In fact, the competent authorities are not always able to detect illegal actions in terms of taxation and protection of workers in real time, and generally do not have the capacity to carry out constant and complete monitoring. The benefits of applying our system can thus be summarized as follows.
In the temporary work contracts based on blockchain and smart contract, employee and business data and agreements between them are analysed automatically in order to facilitate the processing of contracts, make this procedure fully automated, increase the accuracy and processing speeds and allow full compliance with existing contractual contract law.
It can cancel the time to verify the contract's correctness by the competent authorities: if a contract was concluded, the requirements of the employer and the employee were correct. In addition, the competent authority can carry out constant checks in real time by simply accessing data recorded on blockchains. Contracts may also be dispatched automatically to the competent authorities.
The immutability of the data saved on blockchain makes the payments that are consistent with what is stated in the contract terms. Contractual terms must match payment execution and must be based on hours worked.

\section{Conclusion}
The paper proposed the use of blockchain technology to manage temporary employment in order to protect workers' rights and make things easier for undertakings that want to operate in a context of trust and security and fully within the law. We modeled an employment relationship as a state system in which states describe every phase of relationship. We have taken into account all actors of system. One of these actors is the blockchain that has the double role: it is a ledger, public and unchangeable where we the contracts can be registered, and at the same time it has the task of managing, controlling and monitoring the proper execution of employment contract in order to safeguard workers, prevents scams and make the task of businesses and controls by the authorities easier.
In fact, thanks the availability of a decentralized virtual machine, the blockchain is not only a database but also a computing resource.
In order to keep the model clear and simple, in this proposal we have not considered some kind of events that can happen in working life. In facts, in our work we supposed the employment relationship concludes every time in the complete satisfaction of contractual terms and in the time stipulated. Actually, the employment relationship could be concluded differently or end prematurely. For instance, the early conclusion case could occurs when the worker does not reach the workplace, if the employee is made redundant in proper circumstances or when he is not able to complete the assigned job. In future, we want to describe and model those cases with the use of blockchain.

\end{document}